\documentstyle[prl,aps,epsfig]{revtex}
\draft

\begin{document}
\twocolumn[
\hsize\textwidth\columnwidth\hsize\csname@twocolumnfalse\endcsname

\title{The Kondo effect in crossed Luttinger liquids}
\author{Karyn Le Hur}
\address{Theoretische Physik, ETH-H\"onggerberg, 
CH-8093 Z\"urich, Switzerland}
\maketitle
\begin{abstract}
We study the Kondo effect in 
two {\it crossed} Luttinger liquids, using Boundary Conformal Field Theory. 
We predict two types of critical
behaviors: either a two-channel Kondo fixed point with a nonuniversal Wilson
ratio, or a new theory with an anomalous response identical to that
found by Furusaki and Nagaosa (for the Kondo effect in a single
Luttinger liquid). Moreover, 
we discuss the relevance of perturbations like
channel anisotropy in restoring a Fermi-liquid-like 
Kondo fixed point, and we make links with
the Kondo effect
in a two-band Hubbard system modeled by a channel-dependent Luttinger 
Hamiltonian. The suppression 
of backscattering off the impurity produces a model similar to
the four-channel Kondo theory. Consequences are discussed.
\end{abstract}

\vfill
 \twocolumn
\vskip.5pc ]
\narrowtext

\section{Introduction}

The one dimensional (1D) conductors differ
fundamentally from those in three dimensions, where the low-energy properties
can be described very well by Landau's Fermi liquid theory. In 1D, 
the resulting
state is often of the Luttinger liquid (LL) type\cite{one,deux}. 
The physics of 
such low-dimensional systems has received much attention
lately, mainly due to advances in nanofabrication\cite{deuxb} and the 
discovery of
novel 1D materials such as carbon nanotubes\cite{trois}. 
The study of magnetic
impurities in 1D unconventional correlated hosts has attracted
great interest in the last few years. The Kondo effect in a LL
yields two possible fixed points\cite{quatre,Durga,cinq}. 
{\it Either} the system 
behaves rather like a
Fermi liquid (with a nonuniversal Wilson 
ratio and triplet spin quasiparticles\cite{cinq}) {\it or} it
indeed has the non-Fermi-liquid properties predicted by Furusaki and 
Nagaosa\cite{six}.

In this paper, we study the Kondo effect in two {\it crossed} Luttinger 
liquids\cite{seven}, i.e. two correlated 1D metals coupled in a pointlike 
manner via a magnetic 
impurity. An important question is examined:
are the two 
fixed points cited above stable when several conducting
channels interact through a pointlike Kondo coupling? 
The geometry of our system is shown
in Fig. 1. The authors of
ref.\cite{cinq0} have studied the Kondo effect in a 
two-band Hubbard chain 
modeled by a channel-dependent Luttinger
Hamiltonian. On the other hand, for the most general two-band problem 
investigated in ref.\cite{eight}, 
a prominent repulsive Hubbard 
interaction normally destroys the LL phase producing
a metallic spin-gapped
phase with a leading d-wave order
parameter. The resulting Kondo problem becomes very
difficult to handle.

In our case, the two Luttinger
liquids are supposed to
be {\it non-interacting} [except at the impurity site].
In particular, we do not include an electron-electron interaction
for two particles that belong to different conducting channels.
Further experiments on magnetic impurities implanted 
in 1D quantum wires or carbon nanotubes\cite{seven} 
could provide impetus for studying this model.

\begin{figure}
\centerline{\epsfig{file=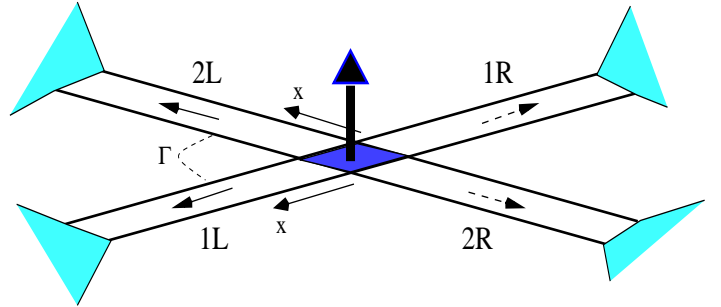,angle=0.0,height=4.1cm,width=9.3cm}}
\vskip 0.3cm
\caption{Two Luttinger liquids coupled only at x=0 via
the Kondo effect. The angle $\Gamma$ is assumed to remain finite.}
\end{figure}

\section{Model}

As long as the angle $\Gamma$ [that is depicted in Fig.1] does not tend
to zero, we can separate the two degenerate Luttinger liquids and
can neglect the electron-electron interaction between channels with
different $i$ ($i=1,2$). As in ref.\cite{seven}, we consider that $x$
measures deviations from the magnetic impurity
in both conducting channels. In such
sense, we have only one coordinate left.
\vskip 0.1cm
The Hamiltonian 
\begin{equation}
{\cal H}={\cal H}_o+{\cal H}_U+{\cal H}_K
\end{equation}
for this two-channel Kondo model
[with left (L) and 
right (R) moving electrons per channel]
consists of the term
for free electrons:
\begin{equation}
{\cal H}_o=v_F\hbox{{\Large \{}}\psi_{iR\sigma}^{\dag}
i\frac{d}{dx}\psi_{iR\sigma}-
\psi_{iL\sigma}^{\dag}i\frac{d}{dx}\psi_{iL\sigma}\hbox{{\Large \}}},
\end{equation}
with $v_F$ being the Fermi velocity and
$i=1,2$ channel index; an electron-electron (e-e) interaction term:
\begin{equation}
{\cal H}_U=U\ j^i_{p} j^i_{p'},\ j^i_{L(R)}=:\psi_{iL(R)\alpha}^{\dag}
\psi_{iL(R)\alpha}:,
\end{equation}
with $U>0$\cite{nos}; and forward and backward scatterings off the impurity:
\begin{eqnarray}
\label{three}
{\cal H}_K&=& \lambda_F\ 
\psi_{iL(R)\alpha}^{\dag}(0){\pmb{$\sigma$}}_{\alpha\beta}\psi_{iL(R)\beta}(0)
\cdot{\bf S}\\ \nonumber
&+& \lambda_B\ 
\psi_{iL(R)\alpha}^{\dag}(0){\pmb{$\sigma$}}_{\alpha\beta}\psi_{iR(L)\beta}
(0)\cdot{\bf S},
\end{eqnarray}
where ${\pmb{$\sigma$}}$ are the usual spin-1/2 matrices.
For the physically relevant case, we have
$\lambda_F=\lambda_B=\lambda_K$ ({\it the
usual Kondo interaction}). 

Conduction electrons of one liquid respond to
a spin flip of the impurity caused by the interactions with electrons
of the other liquid. In this way, there is an {\it induced} interaction between
the liquids. We could also include another interaction of the 
form:
\begin{equation}
\label{four}
\lambda_{m}\epsilon_{ij}[\psi_{iL(R)\alpha}^
{\dag}{\pmb{$\sigma$}}_{\alpha\beta}\psi_{jL(R)\beta}+
\psi_{iL(R)\alpha}^{\dag}{\pmb{$\sigma$}}_{\alpha\beta}\psi_{jR(L)\beta}]{\bf
  S},
\end{equation}
where $\epsilon_{ij}=1$ for $i=1$,$j=2$
and zero otherwise. First, we neglect
the term in Eq.(\ref{four}).

We subsequently 
study this problem by using Boundary Conformal Field Theory (BCFT). The 
heart of the method, 
pioneered by Affleck and Ludwig\cite{neuf,dix}, is to replace the impurity by
a scale invariant boundary condition. It was successfully
applied to study the low-temperature properties of a spin-1/2 magnetic
impurity coupled to a LL\cite{quatre,Durga,cinq}, and to solve the 
Kondo effect in the particular two-band Hubbard chain 
of ref.\cite{cinq0}. 
\vskip 0.2cm
Below, we shall 
precisely discuss how the
geometry of Fig. 1 influences
the two fixed points found in ref.\cite{cinq0}.

\section{Only forward scattering off the impurity}

We first 
study the case of 
only {\it forward} 
scattering off the impurity, i.e. $\lambda_B=0$. 
Let us start with a {\it free} electron gas where $U=0$.

\subsection{Four-channel Kondo model with free electrons}

To solve this case using BCFT, it is convenient to define right and 
left
movers on the half-plane $x\geq 0$ (see Fig.1), so that
\begin{equation}
\psi_{iR\alpha}(t,x)\equiv
\psi_{iL\alpha}(t,-x),
\end{equation}
 with $i=1,2$, and to confine the system to the finite
interval $x\in [-l,l]$.
Fields are {\it left} movers only and it
is useful to rename $\chi_{1\alpha}(x)=\psi_{1L\alpha}(x)$,
$\chi_{2\alpha}(x)=\psi_{1L\alpha}(-x)$, $\chi_{3\alpha}(x)=
\psi_{2L\alpha}(x)$, and
$\chi_{4\alpha}(x)=\psi_{2L\alpha}(-x)$. 
Keeping only $\lambda_F\not=0$ (forward scattering off the
impurity) in (3), it follows:
\begin{equation}
\label{for}
{\cal H}_F=\lambda_F{\bf J}(0)\cdot {\bf S}.
\end{equation}
Here, ${\bf J}$ is the electron spin current density:
${\bf J}(x)=\sum_{i=1}^{k}
\chi_{i\alpha}^{\dag}(x){\pmb{$\sigma$}}_{\alpha\beta}\chi_{i\beta}(x)$
and $k=4$. Note that the
information about the number of channels is contained in the commutation
rules satisfied by these currents\cite{dix},
indicating that $J^a(x)$ form an $SU(2)_{k}$ Kac-Moody algebra. 

Generally, we
must also introduce,
\begin{equation}
{J}(x)=\sum_{i=1}^{k}
\chi_{i\alpha}^{\dag}(x)\chi_{i\alpha}(x), J^A(x)=\sum_{ij\alpha}
\chi_{i\alpha}^{\dag}(x){\bf T}^A_{ij}\chi_{j\alpha}(x),
\end{equation}
where ${\bf T}^A_{ij}$ are the generators of the $SU(k)$ group. Thus, the free
Hamiltonian ${\cal H}_o$ can be rewritten in a suitable Sugawara form,
\begin{equation}
\hskip -0.2cm{\cal H}_o=\frac{v_F}{2\pi}\int dx\frac{J(x)J(x)}{4k}
+\frac{{\bf J}(x){\bf J}(x)}{k+2}+\frac{J^A(x)J^A(x)}{k+2}\cdot
\end{equation}
This allows one to formulate the problem entirely in terms of the electron
spin current, ${\bf J}(x)$. It leads to an effective four-channel 
(left-handed) Kondo theory\cite{dix}. Briefly, we
summarize the arguments below.

The unperturbed problem organizes into a product of {\it three} conformal
towers labeled by the quantum numbers $(Q,j,j_f)$, respectively the charge,
the spin, and the flavor of the system. Starting with an even number of
particles the high-temperature physics is described by the set
$(Q=0,j=0,\hbox{flavor singlet})$. For
the special 
value:
\begin{equation}
\lambda_F^*=\frac{v_F}{k+2}, 
\end{equation}
[the unique solvable point in the isotropic
region, which is commonly identified as the fixed
point of the model\cite{BA}], we can absorb the impurity spin
by redefining the spin current as that of electrons and impurity:
\begin{equation}
{\bf J}(x)
\rightarrow {\bf J}(x)+2\pi{\bf S}\delta(x). 
\end{equation}
For the overscreening
Kondo effect, the absorption of the impurity spin takes place
in the weak-coupling limit and then the groundstate degeneracy g is not
exactly 
1 as in the completely screened situation\cite{dix}, but it takes a {\it
non-integer}
value smaller than 2 (the groundstate degeneracy at 
high temperatures). Then, 
some extra {\it nonmagnetic} degrees of freedom occur
at the impurity site.

Near the fixed point, the Hamiltonian can
be written as the fixed point Hamiltonian plus possible perturbations:
\begin{equation}
{\cal H}={\cal H}_F+\sum_i \gamma_i {\cal O}_i(0).
\end{equation}
We can classify all the possible perturbations ${\cal O}_i$ in the physical
problem according to the representation theory of the underlying Kac-Moody
algebra at the fixed point.

For the overscreening case, non-trivial
boundary operators may appear which do not occur in the bulk
theory. The triplet operator
${\pmb{$\Phi$}}$ always occurs\cite{dix}. 
This selection rule describes a new content of
boundary scaling operators. 
The low-temperature properties are now governed
by the {\it leading-correction-to-scaling boundary operator} (LCBO). This 
must preserve all the symmetries of ${\cal H}_o+{\cal H}_F^*$. We obtain
 a {\it
unique} LCBO: ${\bf J}^{-1}\cdot{\pmb{$\Phi$}}\delta(x)$, which has the 
scaling dimension
$\Delta_S=1+2/(k+2)$ for a left-handed theory. Then, adding 
\begin{equation}
\delta H=\gamma_1{\bf J}^{-1}\cdot{\pmb{$\Phi$}}(0),
\end{equation}
 to
the total Hamiltonian, the leading contribution to low-temperature
thermodynamics is second order in $\gamma_1$. For $k=4$, we have\cite{dix}:
\begin{equation}
C_{imp}\sim T^{2/3}+...,\ \chi_{imp}\sim T^{-1/3}+...\ T\rightarrow 0.
\end{equation}
As pointed out by Fabrizio and Gogolin, the same conclusion 
holds at a particular anisotropic
Kondo limit, namely the so-called Toulouse point\cite{douze}.

\subsection{Role of repulsive interactions in each channel}

When $U\neq 0$, the bulk Hamiltonian
${\cal H}_{TL}$ can also be written on a Sugawara form, using the redefinitions
\cite{quatre}
\begin{eqnarray}
\label{spin}
J_{L(R)}^i(x)&=&\cosh\eta
:\psi_{iL(R)\alpha}^{\dag}(x)\psi_{iL(R)\alpha}(x):\\ \nonumber
&+&\sinh\eta
:\psi_{iR(L)\alpha}^{\dag}(x)\psi_{iR(L)\alpha}(x):\\ \nonumber
{\bf J}_{L(R)}^i(x)&=&:\psi_{iL(R)\alpha}^{\dag}(x)
{\pmb{$\sigma$}}_{\alpha\beta}\psi_{iL(R)\beta}(x):,
\end{eqnarray}
where the currents $J_p^i(x)$ and ${\bf J}_p^i(x)$ 
(where $i=1,2$ and $p=L,R$) satisfy the U(1) and (level-1)
$SU(2)_{1}$ Kac-Moody algebras, respectively and
\begin{equation}
\tanh 2\eta=U/(v_F+U).
\end{equation}
They generate the critical Luttinger bulk Hamiltonian:
\begin{equation}
\label{onze}
{\cal H}_{TL}=\int_0^l dx\frac{v_c}{8\pi}:J_p^i(x)J_p^i(x):
+\frac{v_F}{8\pi}:{\bf J}_p^i(x){\bf J}_p^i(x):.
\end{equation}
Note that ${\cal H}_{TL}$ is invariant under the chiral symmetry
${\cal G}=\{U(1)_L\times U(1)_R\times SU(2)_{1,L}\times 
SU(2)_{1,R}\}^2$. The model yields separation of spin and charge
and the velocity for charge zero sound modes is given by:
\begin{equation} 
v_c=v_F\sqrt{1+2U/v_F}=v_F K^{-1}. 
\end{equation}
The parameter $K=e^{-2\eta}$ can be
identified as the usual Luttinger exponent. At high temperatures,
the spin quasiparticles,
from the SU(2), level-1 Wess-Zumino-Witten conformal field theory, are the
usual spin-1/2 doublets namely {\it spinons} [which bring fractional spins]. 

By analytic continuation, the theory in (\ref{onze}) is equivalent to
a chiral (left-handed) theory on $[-l,l]$.
As the four
currents are coupled via ${\bf S}$, the forward Kondo exchange breaks 
$\{SU(2)_{1,L}\times  SU(2)_{1,L}\}^2$ of ${\cal G}$ down to
the diagonal level-4 subalgebra $SU(2)_4$.
 For conformal theories with an $SU(2)_{k}$ symmetry, the
free energy is proportional to the `central charge' defined as\cite{fe}:
\begin{equation}
C=\frac{3k}{k+2}\cdot
\end{equation}
Thus, we can decompose a $4\times SU(2)_1$ Sugawara
Hamiltonian (with $C=4$) onto an $SU(2)_4$ one (with $C=2$) and a remainder 
describing the flavor sector
(here, an $SU(2)_4$ critical theory with $C=2$, as well). 
This analysis can be routed via the so-called {\it coset} construction
\cite{GKO}.
Then, since only the spin sector $SU(2)_4$ is coupled to the impurity,  
we predict the
same {\it unique} LCBO as for the case without electron-electron interaction
[a boundary operator coming from the charge sector
or the flavor one (only) is characterized by a coupling constant which
 goes to zero when the ultraviolet cutoff goes to infinity]. 
Using the general formula of ref.\cite{cinq}, we obtain a Wilson ratio
\begin{equation}
R_W=\frac{\chi_{imp}C}{\chi C_{imp}}=4(1+K),
\end{equation}
where, $C$ and $\chi$ are the bulk quantities. It should be noted that $R_W$
is universal only for a perfect isotropic
Kondo exchange\cite{Ye} and in the limit
$U\rightarrow 0$: it takes the value $R_W=8$\cite{dix}. 

To conclude, the presence of the
electron-electron interaction makes the Kondo crossover highly 
non-universal. The impurity screening
leads to a new symmetry 
for the bulk Hamiltonian, and then to new N-body excitations
in the infra-red limit [coming from the $SU(2)_4\times SU(2)_4$
(flavor-spin) sectors]. 
However, note that charge quasiparticles with charges $Q=\pm e$,
are still
the usual `holons' of the LL.
\vskip 0.1cm
On the other hand, the low-temperature thermodynamics 
due to the impurity screening is still the same as the one
found in the noninteracting case, because the impurity spin couples only to
{\it individual} electrons. 

\section{Backscattering effects}

Let us now include 
$\lambda_B=\lambda_F\not=0$. First, to confirm that the presence of
backward scattering off the impurity leads to a {\it new} fixed point, we 
start
with $U=0$. With no e-e 
interaction, it is convenient 
to use the so-called Weyl
basis\cite{cinq0}: 
\begin{equation}
\psi^i_{\pm}(x)=[\psi_{L,\sigma}^i(x)\pm
\psi_{R,\sigma}^i(-x)]/\sqrt{2}. 
\end{equation}
Then, $({\cal H}_o+{\cal H}_K)$ transforms 
into a four-channel Kondo theory, but with the impurity coupled to the 
electrons in only
the two positive parity channels, namely $\psi_+^1$ and $\psi_+^2$. 
Thus, we obtain an effective two-channel\cite{Nozieres}
(left-handed) Kondo Hamiltonian\cite{cinq0}. 

Here, it 
is well-known that 
the forward Kondo scattering
term breaks the $SU(2)_{1}\times SU(2)_{1}$ subgroup of 
${\cal H}_o$ down to
$SU(2)_{2}\times {\cal Z}_2$, where ${\cal Z}_2$ is a critical theory with
a central charge C=1/2 equivalent to an {\it Ising} model
\cite{quatre,cinq}.  
The model renormalizes to a marginal non-Fermi liquid with {\it logarithmic} 
corrections. It
can be simply obtained by taking $\gamma_1$ as the unique LCBO (note that
$\Delta_S=3/2$ for $k=2$). The low-temperature thermodynamics
 at the impurity site
is given by\cite{dix}:
\begin{equation}
\label{do}
C_{imp}\propto T\ln(\frac{T_K}{T})+...,\ \chi_{imp}\propto
\ln(\frac{T_K}{T})+...\ T\rightarrow 0
\end{equation}
When $U\neq 0$, the e-e interaction mixes left- and 
right-moving fields, and hence
becomes highly {\it non-local} in the Weyl-basis. Although 
efforts have been
made to handle consistently the non-local terms appearing from 
the interaction\cite{Liu}, in our problem it is very difficult to describe the
Kondo fixed point in the $(\psi_+^1,\psi_+^2)$ basis.
\vskip 0.1cm
However, demanding that any associated LCBO must 
correctly reproduce the
noninteracting limit as $U\rightarrow 0$, the possible critical theories 
can be
deduced:

(A).--- From the spin sector only LCBO with the scaling 
dimension $\Delta_S=3/2$ can occur. The only contribution from the $SU(2)_4$
sector is then the identity and its descendants. {\it This implies
a recombination of conformal towers in the spin sector}.

(B).--- A LCBO including a charge or a flavor field 
unambiguously must be
characterized by a scaling dimension $\Delta_T\rightarrow 1$ as 
$U\rightarrow 0$ [producing no boundary correction in the noninteracting
limit $U\rightarrow 0$].

\subsection{Two-channel Kondo physics when $U\neq 0$}

To guess the precise symmetry of the Hamiltonian in the critical region, we
can use the following points.

First, we can exploit the expectation that the full
Kondo interaction ${\cal H}_K$ can be described as a {\it 
renormalized} boundary condition (selection rule)
on ${\cal H}_{TL}$, analogous to the forward interaction obtained for
$U=0$. In particular, $\lambda_F$
should scale towards the {\it solvable} point $\lambda_F^*=v_F/4$ 
(with k=2) although
$\lambda_B$ goes to strong couplings when $U\neq 0$ or $K\neq 1$
[see below]. Second, the full Hamiltonian must also contain an Ising 
sector. 

As an important consequence, when $U\neq 0$
we must write the fixed-point Hamiltonian ${\cal H}_0+{\cal H}_F^*$ as
 a C=2 critical theory 
with L- and R-movers having an $SU(2)_{2,L}\times SU(2)_{2,R}\times{\cal Z}_2$
symmetry. The presence of backward scattering off the impurity breaks
$SU(2)_{2,L}\times SU(2)_{2,R}$ down to $SU(2)_2$. 

Let us now precisely describe the content of scaling boundary operators.
If ${\bf J}_{L(R)}^1$ and ${\bf J}_{L(R)}^2$ [given
by Eq. (\ref{spin})]
satisfy the level-1 Kac-Moody algebra, then the diagonal currents given
by:
\begin{equation}
{\bf J}_{L(R)}={\bf J}_{L(R)}^1+{\bf J}_{L(R)}^2,
\end{equation}
satisfy the level-2
one. Thus, we can write the Hamiltonian as a sum of an $SU(2)_{2,L}\times
SU(2)_{2,R}$ Sugawara Hamiltonian and an Ising model. Such procedure, for
example, has
been successfully applied to treat the two-leg spin ladder problem\cite{To}.
We can easily complete the `square' at the solvable point $\lambda_F^*=v_F/4$ 
via
the use of the transformation: ${\bf J}_{L(R)}(x)
\rightarrow {\bf J}_{L(R)}(x)+2\pi{\bf S}\delta(x)$. The Kac-Moody algebras
for channels L and R are no longer independent. As for the noninteracting
$U$-limit, $\lambda_F^*=v_F/4$ will be identified as the true fixed point
of the model. However, it should be
noted that the recursion law for $\lambda_F$:
\begin{equation}
\frac{d\lambda_F}{d\ln L}=\frac{{\lambda_F}^2}{2\pi v_F}+
\frac{{\lambda_B}^2}{2\pi v_F}-\frac{k}{2}\frac{{\lambda_F}^3}
{(2\pi v_F)^2}+...,
\end{equation}
does not allow to find the precise forward Kondo exchange 
infra-red value, namely $\lambda_F^*$. We can only assume 
that, as for $U=0$, the presence
of the last term which occurs with a {\it minus} sign should
prevent $\lambda_F$ to flow to strong couplings.

The eigenstates in
the $SU(2)_{2,L}\times SU(2)_{2,R}$ sector appear in conformal towers labeled
by the spin quantum numbers $j=0,\ 1/2,\ 1$. The corresponding primary fields
are the identity ${\bf 1}$, the fundamental field $g$, and the triplet 
operator (a 3$\times$3 matrix) $\pmb{$\Phi$}=\sum_{i,j}{\Phi}_{Li}{\Phi}_{Rj}$.
They have the scaling 
dimensions, $\Delta_S=\frac{1}{2}j(j+1)$. Similarly, there are three primary
fields in the Ising sector given by $\phi={\bf 1},\ \sigma,\
\epsilon$ with scaling dimensions $\Delta_{I}$ given by $0,\
\frac{1}{8},\ 1$, respectively. 

In respect to the noninteracting case
$U\rightarrow 0$, the absorption of the impurity must give for
forward scattering $(j,\phi)=(0\ \hbox{or}\ 1,{\bf 1})$
\cite{dix}. Simply, through the
examination of spin singlets from $SU(2)_{2,L}\times
SU(2)_{2,R}$, one obtains the following LCBO:
\begin{equation} 
\delta{\cal H}=\gamma_1\hbox{\Large\{}{\bf J}_L^{-1} 
\pmb{$\Phi$}_{L}(0)+{\bf J}_R^{-1}\pmb{$\Phi$}_{R}(0)\hbox{\Large\}}. 
\end{equation}
By construction, $\pmb{$\Phi$}_{L}$ and $\pmb{$\Phi$}_{R}$ have the halved 
dimension $1/2$. Thus, we can easily
check that $\delta{\cal H}$ produces a two-channel-like 
Kondo fixed point with transport properties given in Eq. (\ref{do}). 
We like to point out the following remark.
Although
${\bf J}_L$ and ${\bf J}_R$ are coupled through the impurity screening, the
symmetry $SU(2)_{2,L}\times
SU(2)_{2,R}$ of the total Hamiltonian cannot be broken at the fixed
point because a descendant field ${\bf J}_p^{-1}$
with $p=L,R$ acts only on a primary field from the p-sector. 
The same fixed point has been found for the 
overscreened Kondo effect in a two-band Hubbard chain\cite{cinq0}, {\it meaning
that at low temperatures the geometry of the system does not affect spin
properties near the impurity}. 
We can check that the Wilson ratio is universal only when
$U\rightarrow 0$; it takes the value 8/3\cite{dix}.

\subsection{Generalized tunneling process \`a la Furusaki-Nagaosa}

Neglecting the $\lambda_m$ term, for 
{\it each} LL the charge eigenstates organize
into a product of two U(1) conformal towers, labeled by the two quantum
numbers $(Q_i,\Delta Q_i)$, the sum and difference of net charge in the left 
and right channels. Introducing the usual 
charge variables $(i=1,2)$,
\begin{equation}
(J_L^i+J_R^i)=\frac{1}{\sqrt{2\pi K}}\partial_x\phi_{ci},\ 
(J_L^i-J_R^i)=\sqrt{\frac{K}{2\pi}}\Pi_{ci},
\end{equation}
the charge part of the free Hamiltonian can be identified as two
independent Luttinger models\cite{one,deux}.

Now, we must carefully treat backward
scattering off the impurity. Indeed, the corresponding
term in (\ref{three}) breaks the chiral U(1) invariance of 
${\cal H}_{TL}$. The selection rule for combining the two U(1)
conformal towers may change. Thus, $\Delta Q_i$ is no longer
restricted to zero, and the charge sector should make nontrivial contributions
to the content of scaling operators leading to another possible fixed
point in the critical region.

The backscattering term (\ref{three}) is usually expressed in the so-called
spinon basis as\cite{Durga,cinq}
\begin{equation}
\label{uu}
{\cal H}_B=\lambda_B\sum_{i=1,2}\hbox{\Large\{}
\hbox{Tr}(g_i{\pmb{$\sigma$}})
\cos\sqrt{2\pi}\phi_{ci}(0)\hbox{\Large\}}\cdot{\bf S},
\end{equation}
and the spin operators $g_i\in SU(2)_{1,L}\times SU(2)_{1,R}$. Using simple
scaling arguments (with $L\equiv 1/T$):
\begin{equation}
\frac{d\lambda_B}{d\ln L}=\frac{1}{2}(1-K)\lambda_B+{\cal
  O}(\frac{\lambda_B\lambda_F}{\pi v_F}),
\end{equation}
we find that prominent backscattering
off the impurity supports a Kondo effect for ferromagnetic as well as
antiferromagnetic Kondo exchanges\cite{Vo}.
The Kondo 
temperature yields the same power-law dependence on the exchange coupling
$T_K\propto {\lambda_B}^{2/(1-K)}$ as for the single LL
case\cite{six,Lee}.
To summarize, when $K\neq 1$ the flow of $\lambda_B\neq 0$
goes to infinity whereas the forward Kondo scattering exchange scales
to the precise intermediate value given by Eq. (8), with k=2.

When $T\ll T_K$, we have the formation of a 
bound state (with spin S=0)
between any electron near the Fermi level and the impurity
spin. However, a nonmagnetic
 extra degree of freedom remains at the impurity
site because $\lambda_F^*$ is not too strong [let us remind that 
only the forward Kondo exchange can really absorb or screen
the impurity spin]. Precisely, 
for $k=2$, the groundstate
degeneracy is exactly $g=\sqrt{2}$\cite{dix}, and it 
can be interpreted as a residual
Majorana fermion at the origin\cite{EK}. 

On the other hand, the fact that $\lambda_B\rightarrow +\infty$ can be
interpreted as follows. In the infra-red region, 
the cosine terms of Eq. (\ref{uu}) become
pinned at the
origin and $\langle \cos\sqrt{2\pi}\phi_{ci}(0)\rangle=$
constant or $\phi_{ci}(0)=\sqrt{\pi/2}$\cite{cinq}.
Simply, it means that the charge quasiparticles ({\it holons})
move completely away from the origin
[despite the relatively weak value
of the forward Kondo exchange at the fixed point], due to the concrete 
spin-charge separation occurring in a
1D metallic wire for $K\neq 1$: only spin degrees of freedom couple to
the impurity in the infra-red region.
Finally, since a bound state between an electron of the Fermi sea and the
impurity spin acts as a strong nonmagnetic 
barrier at $x=0$ and since
$\lambda_B\rightarrow +\infty$, exotic tunneling phenomena can take place.
In the infra-red limit, we must
decompose the backscattering term $\lambda_B$ (written via $g_1$ and $g_2$) 
in the (Ising)$\otimes$g basis (which
has been used to absorb the impurity spin). After some complicated
algebra, the result is\cite{To}:
\begin{equation}
\hbox{Tr}(g_1{\pmb{$\sigma$}})+\hbox{Tr}(g_2{\pmb{$\sigma$}})=\sqrt{2}
\hbox{Tr}(g{\pmb{$\sigma$}})\cdot\sigma.
\end{equation}
The lowest dimension operator with
$\Delta Q_i\neq 0$ allowed by the forward selection rule is obtained from
$(Q_i,\Delta Q_i,j,\phi)=(0,\pm 2,0,{\bf 1})$, has the scaling dimension
$1/2K$ and can be written as: $\cos\sqrt{2\pi}\tilde{\phi}_{ci}(0)$. 
Then, possible couplings of $SU(2)_2$ and Ising towers to the U(1)
towers yield the following candidate LCBO:
\begin{equation}
\delta {\cal
  H}=\gamma_2\hbox{Tr}(g{\pmb{$\sigma$}})\cdot\sigma\sum_{i=1,2}
\cos\sqrt{2\pi}\tilde{\phi}_{ci}(0),
\end{equation}
and $\gamma_2\propto 1/\lambda_B$.
Such term describes a collective tunneling process of two electrons (one in
each LL), which breaks the spin singlet at the impurity site.

Since there is no Hubbard coupling between  
channels 1 and 2, a tunneling phenomenon including a renormalized
(channel-dependent) LL charge parameter 
cannot occur. This is the main difference with the 
Kondo effect in a two-band Hubbard chain\cite{cinq0}.
Here, physical properties 
exhibit an exact duality between high- and low-temperature fixed
points, replacing $K\rightarrow 1/K$\cite{cinq}.
We can check that such an operator with scaling dimension
$\Delta_T=\frac{1}{2}(\frac{1}{K}+1)$ [which goes to 1 as U goes to zero]
shows the same anomalous
scaling in temperature
 as the one predicted by Furusaki and Nagaosa for the Kondo effect
in a LL\cite{six}. Thus, the impurity specific heat and the 
conductance also exhibit the same anomalous temperature dependence with 
a leading term (at $T\rightarrow 0$):
\begin{equation}
G_{imp}(T)\propto T^{(1/K)-1},\
C_{imp}(T)\propto T^{(1/K)-1},
\end{equation}
which
vanishes when $K\rightarrow 1$\cite{quatre,cinq,six}
i.e. for the noninteracting case. The current-voltage
curve associated with this tunneling process obeys: 
\begin{equation}
G(V)\equiv \frac{dI}{dV}
\propto |V|^{(1/K)-1},
\end{equation}
[thermal energy has been replaced by electric energy].
When $K=1$ a linear I-V curve is predicted, consistent with expectations
for non-interacting electrons which are partially transmitted through a
nonmagnetic barrier. For $K\neq 1$, we obtain
\begin{equation}
I\propto |V|^{1/K},
\end{equation}
 and then the linear conductance is strictly zero. This is
a simple reflection of the suppressed density of states in a LL.

\section{Conclusions and discussions on relevant perturbations}

Summarizing, we have studied the low-temperature properties of a spin-1/2
magnetic impurity coupled to two crossed 
conducting channels, each described by a Luttinger
model. Using Boundary Conformal Field Theory, we have reached the
important conclusion that the problem still
admits two possible fixed points: either
the theory remains a marginal non-Fermi liquid with logarithmic corrections
in the presence of electron-electron interactions, or electron correlations
drive the system to another non-Fermi liquid fixed point obtained originally
by Furusaki and Nagaosa for the Kondo effect in a LL.

However, as in the case without
e-e interaction\cite{Fa},
the previous marginal non-Fermi liquid is unstable in presence
of a {\it small} channel anisotropy 
$\delta=(\lambda_F^1-\lambda_F^2)$. Adding the corresponding term
\begin{equation}
{\cal H}_A=\delta({\bf J}_L^1-{\bf J}_L^2+{\bf J}_R^1-{\bf J}_R^2){\bf S}
=\delta(\epsilon_L\pmb{$\Phi$}_{L}+\epsilon_R\pmb{$\Phi$}_{R}){\bf S},
\end{equation}
to the Hamiltonian
destabilizes the symmetric forward scattering fixed point. 
As in the Kondo effect in a LL\cite{cinq} or the famous two-impurity
model in a three dimensional Fermi liquid environment\cite{af}, the LCBO 
${\bf J}^{-1}_p{\pmb{$\Phi$}}_p$ $(p=L,R)$ is excluded by 
parity conservation. We have used the notations: $\epsilon=
\epsilon_L\epsilon_R$. Here, $\epsilon_p$ enters as an allowed boundary
operator of scaling dimension $\Delta_{I}=1$, producing a one-channel
(Fermi-liquid-like) fixed point, ruled by the new selection rule
$\delta^*\rightarrow +\infty$. There are now three irrelevant leading
operators of dimension 2, namely ${\bf J}_{L}^1{\bf J}_{L}^1$, 
${\bf J}^1_R{\bf J}^1_R$, ${\bf J}^1_L{\bf J}^1_R$\cite{cinq}. 
To conclude, either
Fermi-liquid-like [with triplet excitations driven by an
$SU(2)_{k=2}$ CFT] 
or non-Fermi liquid {\it \`a la} Furusaki-Nagaosa
could be still realized experimentally in 
multi-channel 1D quantum wires or
carbon nanotubes
satisfying the geometry presented here. 

Note also that the
suppression of backscattering off the impurity  
produces a low-energy physics identical to that of the four-channel
Kondo model. 

Finally, $\lambda_m=\lambda_F=\lambda_B\neq 0$ seems also to
be a relevant [but not very realistic] perturbation.
Indeed, passing to an odd-even 
parity basis, $(a,b)=1/\sqrt{2}(\psi_1\pm \psi_2)$ [when $U\rightarrow 0$]
the impurity couples only to the fermionic channel $a$. This also
leads to a Fermi-liquid-like
fixed point or to the Furusaki-Nagaosa non-Fermi-liquid one. 

A summary of various physical behaviors is given
in Table 1.

I thank A. Honecker for pointing out to me
ref.\cite{To} and C. Schweigert for discussions on coset constructions.

\pagebreak\onecolumn

\begin{center}
\begin{tabular}{||c|c|c|c|c||}
\tableline
Impurity & Susceptibility & Specific heat & 
Conductance/Resistivity & Fixed point\\
\tableline
$\lambda_F\neq 0$ and $\lambda_B=0$ & $T^{-1/3}$ & $T^{2/3}$ & $\rho\propto {T^{1/3}}^*$ & 4-channel-like\\
\tableline
\hskip 0.2cm $\lambda_F=\lambda_B\neq 0$\hskip 0.9cm {\bf 1-}& const. & $T^{(1/K)-1}$ & $G\propto T^{(1/K)-1}$ & Furusaki-Nagaosa \\
\hskip 3.1cm {\bf 2-}& $\ln T$ & $T\ln T$ & $\rho\propto \sqrt{T}^*$&
2-channel-like, Or\\
If $\delta,\lambda_m\neq 0$ & const.$^*$ & $T^*$ & $\rho\propto {T^2}^*$ & 
1-channel-like\\
\tableline
\end{tabular}
\end{center}

\vskip 0.3cm
{\it Table 1: Different fixed points and physical behaviors reported
in this paper, for the
Kondo effect in crossed Luttinger liquids. 
Notations are explained in the text and 
$^*$ is from ref.\cite{dix}.}


\begin{references}
\bibitem{nos} Spin interactions are rescaled to zero in the LL for $U>0$.
\end{references}
\end{document}